\documentclass[aps,pre,superscriptaddress,,amsmath,amssymb]{revtex4-2}
\usepackage{graphicx}% Include figure files
\usepackage{dcolumn}% Align table columns on decimal point
\usepackage{bm}% bold math
\usepackage{xcolor}

\begin{document}

\renewcommand{\d}[2]{\frac{d #1}{d #2}} % derivative
\newcommand{\dd}[2]{\frac{d^2 #1}{d #2^2}} % second derivative
\newcommand{\pd}[2]{\frac{\partial #1}{\partial #2}}  % partial derivative
\newcommand{\pdd}[2]{\frac{\partial^2 #1}{\partial #2^2}} % second partial derivative
\newcommand{\paren}[1]{\left( #1 \right)} % parenthesis
\newcommand{\cor}[1]{\left[ #1 \right]} % square parenthesis

%\preprint{}

\title{Chemical oscillators synchronized via an active oscillating medium: dynamics and phase approximation model}


\author{David Garc\'{i}a-Selfa}
\affiliation{CRETUS Institute. Group of Nonlinear Physics. Dept. of Physics. Universidade de Santiago de Compostela. 15782 Santiago de Compostela, Spain}
\affiliation{CESGA (Supercomputing Center of Galicia). 15705 Santiago de Compostela, Spain}
\author{Gourab Ghoshal}
\affiliation{Department of Physics \& Astronomy, University of Rochester, Rochester, New York 14607, USA}
\author{Christian Bick}
\affiliation{Centre for Systems Dynamics and Control and Department of Mathematics, University of Exeter, Exeter EX4 4QF, UK}
\author{Juan P\'{e}rez-Mercader}
\affiliation{Department of Earth and Planetary Sciences. Harvard University, Cambridge, MA 02138, USA}
\affiliation{The Santa Fe Institute, 1399 Hyde Park Road, Santa Fe, NM 87501, USA}
\author{Alberto P. Mu\~{n}uzuri}
\affiliation{CRETUS Institute. Group of Nonlinear Physics. Dept. of Physics. Universidade de Santiago de Compostela. 15782 Santiago de Compostela, Spain}

%\thanks{}
%\affiliation{}


%\date{\today}

\begin{abstract}
Different types of synchronization states are found when non-linear chemical oscillators are embedded into an active medium that interconnects the oscillators but also contributes to the system dynamics. Using different theoretical tools, we approach this problem in order to describe the transition between two such synchronized states. Bifurcation and continuation analysis provide a full description of the parameter space. Phase approximation modeling allows the calculation of the oscillator periods and the bifurcation point.
\end{abstract}

\keywords{chemical oscillators, synchronization, bifurcation theory, phase approximation, mobbing state}

\maketitle

\section{Introduction}

Chemical oscillatory behavior is evidence of complex, highly nonlinear dynamics, and is ubiquitous in nature~\cite{Strogatz2003,Murray1993}. In many cases, sub-units exhibiting oscillatory behavior couple together in large assemblies giving rise to a collective behavior of which a particularly important phenomenon is synchronization. Synchronization plays important roles in multiple biological and technical settings, for instance in the synchronized flashing of fireflies~\cite{Buck1988}, in cardiac pace-makers~\cite{Peskin1975,Torre1976,Mirollo1990}, in yeast cells~\cite{Ghosh1971}, the firing of neurons~\cite{Izhikevich2007}, in arrays of Josephson junctions~\cite{Wiesenfeld1998} and semiconductor lasers~\cite{Hohl1997} among numerous other examples.

Given its ubiquity, the mechanism involved in synchronization has been the subject of rigorous study through both analytical considerations (phase models based on the Kuramoto-family of models ~\cite{Acebron2005,Nakao2016}) as well as experimental realizations including coupled electrochemical oscillators and reactors~\cite{Kiss2005}.  Of particular note are populations of catalyst-loaded oscillatory beads. A typical setup consists of a large number of beads in which the oscillatory Belousov--Zhabotinsky~(BZ) reaction takes place~\cite{Taylor2009}. These beads are immersed in a well-stirred active medium which acts as coupling between the population~\cite{Tinsley2010}, with oscillations being triggered by the contact of the beads with the medium. Several interesting dynamical behaviors have been reproduced in this setting including phase synchronization~\cite{Taylor2011}, quorum sensing~\cite{Taylor2009} and amplitude entrainment.

Most studies carried out this far focus on the behavior of the beads while treating the active medium as just a mean to couple them, while ignoring its role as a potential oscillator in itself.  Recently, however, a relatively new synchronization phenomenon was reported in numerical and experimental investigations, where the beads as well as the active medium are driven into a common high amplitude, low-frequency super-synchronized state of oscillations~\cite{Ghoshal2016}. Interestingly, this occurred in the strong coupling limit, where previously the only reported state was that of oscillator death~\cite{Taylor2009}. Beyond an experimental setting, this phenomenon has practical relevance, being similar to exotic states of synchronization as found in Interictal Epileptogenic Discharges, a known neuro-pathology~\cite{Prince1986,deCurtis2001}, and temperature mediated synchronization of the chirping of crickets~\cite{Walker1969}. Qualitative arguments and numerical analysis suggested the presence of higher harmonics~\cite{Ghoshal2016} in the coupling function between the beads and the medium, although the precise forms were never presented. 

In this paper, we fill this gap, by shedding light on the dynamical mechanisms behind this super-synchronized state of oscillations between the beads and the medium. In our theoretical approach, we consider a reduced system with two interacting oscillators, the collection of beads (that are synchronized a priori through standard coupling) and the active medium itself. The active medium  is catalyst free and always coupled to all other oscillators. This reduced approach enables us to uncover the bifurcation structure for the system, across most of its known dynamical states. Reducing the dynamics to a set of phase equations, we calculate for the first time the period of oscillations for this exotic state~\cite{Nakao2016,Pietras2019}. In addition, we present the precise form for the higher-harmonics in the coupling function between the two oscillatory systems.

The paper is organized as follows. In Section~\ref{sec:Dynamic} we introduce the model equations with the needed simplifications and present the obtained state space diagram using bifurcation and continuation analysis (in this paper, we use the term \textit{space state diagram} instead of \textit{phase diagram}, as usually used in dynamical systems theory, in order to avoid confusions with the phase of the oscillators following \cite{Nakao2016}). Finally, in Section ~\ref{sec:PhaseApprox}, we use the phase approximation model to study the transition between the synchronized and the super-synchronized states (mobbing states), calculating interaction functions, the periods of the oscillations in both of the states and the Fourier expansion of the interaction functions where we can see how the Fourier modes change in the transition between the synchronized and the super-synchronized states. The manuscript concludes with a section presenting the conclusions of this work.

This manuscript really points out the importance of an active medium as the means to couple oscillators. In fact, the active connecting medium introduces a great variety of non-trivial behaviors that cannot be described neither understood without its active dynamics. 

\section{\label{sec:Dynamic}Dynamic study: bifurcation and continuation analysis}
We consider a system of ~$n_{\text{beads}}$ coupled chemical oscillators. Each oscillator is a resin bead loaded with the catalyst of the oscillatory BZ reaction. These beads are immersed in a surrounding solution containing all the chemicals of the BZ reaction except for the catalysts that it is in only present on the surface of the beads.
The reactor is a continuously stirred tank \cite{Taylor2009,Ghoshal2016}. 

This system can be described by the following set of differential equations. The dynamics of bead $i\in\{1,...,n_{\text{beads}}\}$ is described with the 3-variable Oregonator model~\cite{Taylor2009,Ghoshal2016} given by
\begin{widetext}
\begin{equation}
\label{eq:EqBeads}
\left\{
\begin{aligned}
\epsilon \pd{x_{i}(t)}{t}&=x_{i}(t) \paren{1-x_{i}(t)}+y_{i}(t)\paren{q-x_{i}(t)}-K_{\text{ex}} \paren{x_{i}(t)-x_{s}(t)} \\
\epsilon'  \pd{y_{i}(t)}{t} &=2h z_{i}(t)-y_{i}(t)\paren{q+x_{i}(t)}-K_{\text{ex}} \paren{y_{i}(t)-y_{s}(t)}\\
\pd{z_{i}(t)}{t} &=x_{i}(t)-z_{i}(t)\\
\end{aligned}
\right.
\end{equation}
\end{widetext}
where $x_i$, $y_i$ and $z_i$ are the dimensionless variables representing the concentrations of activator, inhibitor, and catalyst, respectively for bead~$i$. The quantities~$x_s, y_s$ are the dimensionless concentrations of activator and inhibitor in the surrounding solution (active medium), respectively. Note that the system is well-stirred and the concentration at any location of the surrounding medium is supposed to be the same. The parameter~$K_{\text{ex}}$ is the exchange rate constant between the beads and the surrounding solution. The parameters~$\epsilon$, $\epsilon'$, $q$ and~$h$ are related to reaction rates and initial concentrations~\cite{Ghoshal2016}.

Since the surrounding solution (catalyst-free BZ reaction) interacts with all beads by exchanging activator and inhibitor in the reactor, it formally plays the role of coupling between the beads. For a well-stirred tank reactor, the dynamics of the concentrations of activator and inhibitor in the surrounding solution are given by
\begin{widetext}
\begin{equation}
\label{eq:EqSoln}
\left\{
\begin{aligned}
\epsilon \pd{x_{s}(t)}{t}&=x_{s}(t) \paren{1-x_{s}(t)}+y_{s}(t)\paren{q -x_{s}(t)}+ \frac{\left<V \right>_n}{V_{s}}K_{\text{ex}}\sum_{i=1}^{n_{\text{beads}}} \paren{x_{i}(t)-x_{s}(t)} \\
\epsilon'  \pd{y_{s}(t)}{t} &=-y_{s}(t)\paren{q +x_{s}(t)}+\frac{\left<V\right>_n}{V_{s}}K_{\text{ex}}\sum_{i=1}^{n_{\text{beads}}} \paren{y_{i}(t)-y_{s}(t)} \\
\end{aligned}
\right.
\end{equation} 
\end{widetext}
where the parameter $\left<V\right>_n$ represents the average volume of the beads and the parameter~$V_{s}$ is the total volume of the surrounding solution. Note that although the surrounding solution does not contain catalyst by itself, it does contain the beads that have the catalyst incorporated. Thus, the surrounding solution under these circunstances can potentially exhibit oscillations.

In the following, we will focus in understanding the transition between the synchronized state to the mobbing state, i.e., the transition from the state characterized for all the chemical oscillators oscillating  synchronous to the state in which all the beads oscillate in synchrony and with the surrounding medium also oscillating with the same amplitude and frequency.

As the transitions we are interested in involve that all the beads are already in a synchronized state, we consider that all oscillators, excluding the active medium, are identical and, thus, we can consider the synchronization manifold where the state of all oscillators is equal. Specifically, we assume that $x_i = x_b$, $y_i = y_b$, $z_i = z_b$ for all $i\in\{1, \dotsc, n_{\text{beads}}\}$ (the index~$b$ denotes a bead). Thus, on the synchronization manifold, the model~\eqref{eq:EqBeads},\eqref{eq:EqSoln} reduce to the five-dimensional system
\begin{widetext}
\begin{equation}
\left\{
\begin{aligned}
\epsilon \pd{x_{b}(t)}{t}&=x_{b}(t) \paren{1-x_{b}(t)}+y_{b}(t)\paren{q-x_{b}(t)}-K_{\text{ex}} \paren{x_{b}(t)-x_{s}(t)} \\
\epsilon'  \pd{y_{b}(t)}{t} &=2h z_{b}(t)-y_{b}(t)\paren{q+x_{b}(t)}-K_{\text{ex}} \paren{y_{b}(t)-y_{s}(t)}\\
\pd{z_{b}(t)}{t} &=x_{b}(t)-z_{b}(t) \\
\epsilon \pd{x_{s}(t)}{t}&=x_{s}(t) \paren{1-x_{s}(t)}+y_{s}(t)\paren{q -x_{s}(t)}+ \rho K_{\text{ex}}\paren{x_{b}(t)-x_{s}(t)} \\
\epsilon'  \pd{y_{s}(t)}{t} &=-y_{s}(t)\paren{q +x_{s}(t)}+\rho K_{\text{ex}}\paren{y_{b}(t)-y_{s}(t)}\\
\end{aligned}
\right.
\end{equation}
\end{widetext}
with  $\rho=n_{\text{beads}}\frac{\left<V\right>_n}{V_{s}}$ is the density of the system.

\subsection*{State space diagram}
\begin{figure}[hpt]
 \centering
 \includegraphics[width=0.95\textwidth]{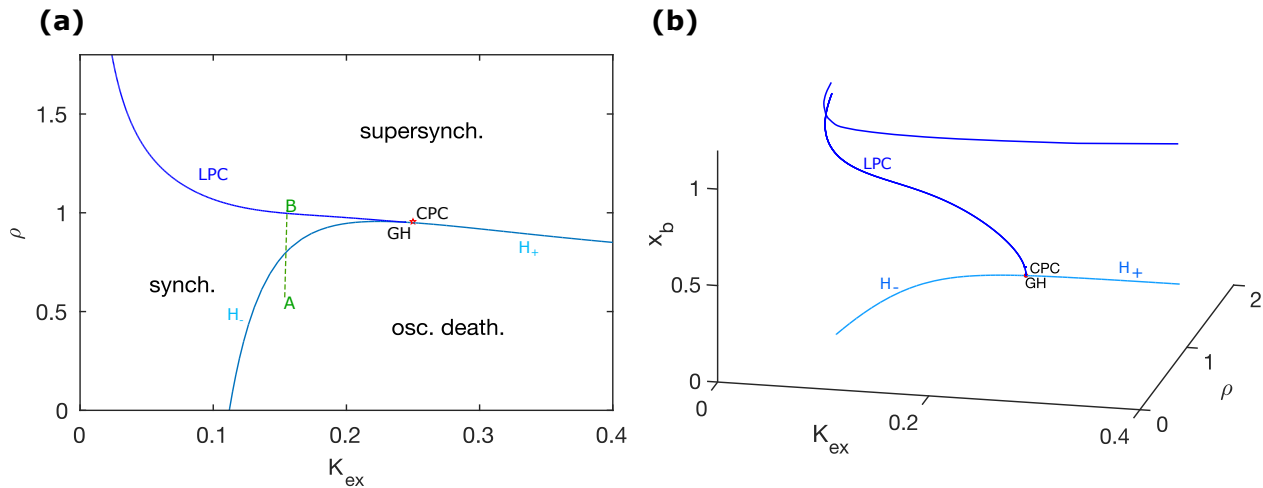}
   \caption{State space diagram. (a) Different behaviors observed in the system when $K_{ex}$ and $\rho$ are varied. (b) Same state space diagram but in a three-dimensional perspective with the value os $x_b$ at the stationary is plotted in the vertical axis.}
 \label{Figure 1}
\end{figure}

\begin{figure}
 \includegraphics[width=0.90\textwidth]{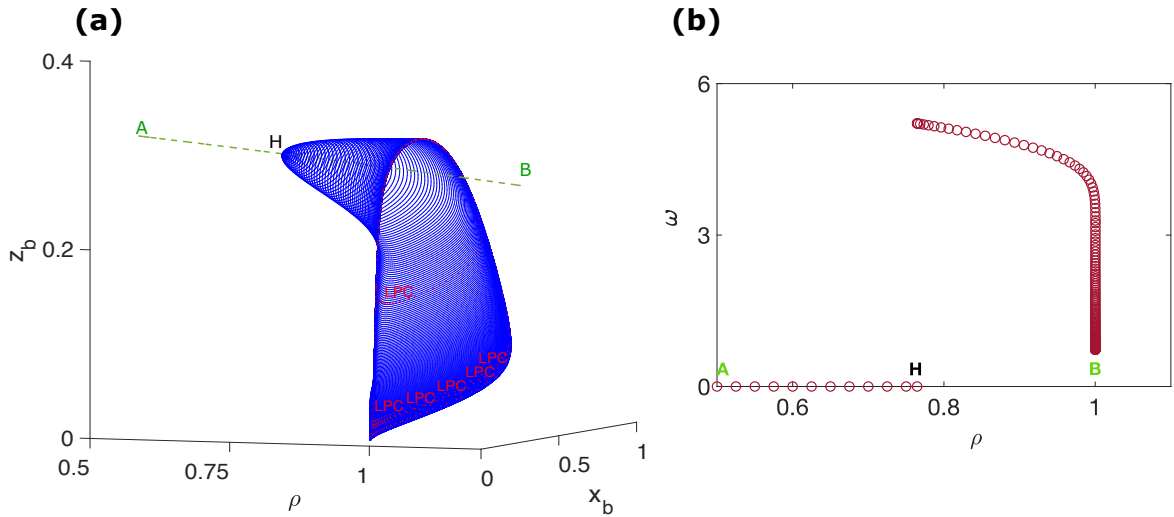}
   \caption{\textbf{(a)} Limit cycles corresponding to the transitions from equilibrium point A (oscillation death) in point A to super-synchronized state in point B passing through the synchronized state (between the supercritical Hopf bifurcation, $H$,  and the saddle-node bifurcations of periodic orbits, $LPC$). Dashed green line A-B corresponding to that shown in \textbf{Figure~\ref{Figure 1}a}. \textbf{(b)} Frequencies corresponding to these transitions.}
 \label{Figure 2}
\end{figure}

For the five-dimensional simplified system, we obtain the state space diagram using continuation and bifurcation analysis of dynamical systems software (\texttt{Matcont}~\cite{Dhooge2008} and \texttt{AUTO}~\cite{Ermentrout2002}). In \textbf{Figure~\ref{Figure 1}} we show the state space diagrams obtained with the model in Eq. (3). \textbf{Figure~\ref{Figure 1}a} displays all the observed behaviors as a function of the exchange rate constant between the beads and the surrounding medium ($K_{ex}$) and the density of beads ($\rho$). As a first observation it is noteworthy the fact that the same behaviors observed both experimentally and numerically are also observed with the same distribution on the state space diagram \cite{Ghoshal2016}. The same bifurcation diagram is plotted in \textbf{Figure~\ref{Figure 1}b} in a 3D perspective where the vertical axis corresponds with the value of the variable $x_b$ for the beads. This new representation unveils the details of the different bifurcations involved in the transitions analyzed. In both representations we observe, the generalized Hopf bifurcation point ($GH$) separating the two branches of supercritical Hopf bifurcation ($H_-$), where  the first Lyapunov coefficient is negative, and subcritical Hopf bifurcation ($H_+$), where the first Lyapunov coefficient is positive, and the saddle-node bifurcation of periodic orbits ($LPC$ curve), where the system has a unique non-hyperbolic limit cycle with the nontrivial Floquet multiplier $+1$. On the other hand, coincident with the~$GH$ point, we have a cusp point of cycles ($CPC$) as this bifurcation separates the supercritical behavior from the subcritical one. This simplified model captures the dynamics of the system that was described in Figure 2-d of~\cite{Ghoshal2016}. Note that, since all the beads are identical by construction in our system, the non-synchronization phase shown in \cite{Ghoshal2016} does not appear.
In \textbf{Figure~\ref{Figure 2}a} we can see the periodic orbits in the transitions from equilibrium (this state is the equivalent to the oscillations death in the complete ($3\times n_{\text{beads}}+2$)-dimensional model) in point~A to super-synchronized state in point~B passing through the synchronized state (between the supercritical Hopf bifurcation~$H$ and the saddle-node bifurcations of periodic orbits~$LPC$). Note that the rapid change of the limit cycle is a Canard explosion that arises even in the simple Oregonator model~\cite{Peng1991,Brons1991,Krupa2001}. We can also see the drastic decrease in frequency in the Canard explosion in \textbf{Figure~\ref{Figure 2}b}. Thus, the Canard explosion observed in a simple Oregonator can also be observed in the synchronized population of Oregonators coupled via the active medium.

\section{\label{sec:PhaseApprox}Phase approximation model}

In the following, we use phase approximation in order to obtain the periods of the beads in the synchronized and super-synchronized phases. Given an oscillating system, its state is described by its position along its limit cycle (its phase). If we have two uncoupled oscillators, their phases lie on a torus. If the two oscillators have stable limit cycles and they are weakly coupled, the torus persists and we can describe their state by their phases~\cite{Nakao2016,Pietras2019}.

\subsection*{Theoretical model}

Consider two oscillators coupled through an interaction function~$\mathbf{G}_i$ such that the dynamics are given by
\begin{equation}
\label{eq:PhaseRed}
\left\{
\begin{aligned}
\d{\mathbf{X}_1(t)}{t}&=\mathbf{F}(\mathbf{X}_1)+K \mathbf{G}_1(\mathbf{X}_1,\mathbf{X}_2)\\
\d{\mathbf{X}_2(t)}{t}&=\mathbf{F}(\mathbf{X}_2)+K \mathbf{G}_2(\mathbf{X}_1,\mathbf{X}_2)
\end{aligned}
\right.
\end{equation} 
If the coupling strength~$K$ is small, the dynamics can be reduced to a phase description, that is, the state of each oscillator is determined by a phase variable $\theta_i$, $i\in\{1,2\}$ on the circle which evolves according to
\begin{equation}
\left\{
\begin{aligned}
\d{\theta_1(t)}{t}&=\omega_1+K H_1(\theta_2-\theta_1)\\
\d{\theta_2(t)}{t}&=\omega_2+K H_2(\theta_1-\theta_2)
\end{aligned}
\right.
\end{equation}
where~$\omega_i$ are the intrinsic frequencies of the oscillators and~$H_i(\phi)$ are the phase interaction functions (that depend only on the phase difference $\phi=\theta_2-\theta_1$). The phase interaction function is computed by averaging
\begin{widetext}
\begin{equation}
H_i(\phi)=\frac{1}{T} \int_0^T \mathbf{Z}(t)\cdot \mathbf{G}_i(\mathbf{X_0}(t+\phi),\mathbf{X_0}(t)) dt=\frac{1}{2\pi} \int_0^{2\pi} \mathbf{Z}(\varphi)\cdot \mathbf{G}_i(\mathbf{X_0}(\varphi+\phi),\mathbf{X_0}(\varphi)) d\varphi
\end{equation}
\end{widetext}
where~$\mathbf{X_0}(t)$ is stable limit cycle,~$\mathbf{Z}(t)$ is the adjoin or phase response curve (PRC)---the phase shift function obtained when the system that lies on its limit cycle is infinitesimally perturbed---and $\cdot$ denotes the scalar product of vectors. Both~$\mathbf{Z}(\theta)$ and~$H_i(\phi)$ can be obtained numerically using \texttt{XPPAUTO}~\cite{Ermentrout2002}. For a theoretical derivation of the PRC by means of the adjoint method and the average method for calculation of the interaction function, see~\cite{Nakao2016}.

\subsection*{Application to our problem}

The problem considered consists on a bead (all the initial beads, with density $\rho$, are considered to be completely synchronized among them) uncoupled to the rest of the system (the surrounding solution). Note that we are assuming that all the beads are already in a synchronized state and, thus, can be represented by one single set of equations as in previous section. This system will be numerically solved using the above mentioned software. The equations describing two identical copies of the system are
\begin{widetext}
\begin{equation}
\label{eq:Uncoupled}
\left\{
\begin{aligned}
\pd{x_{b_i}(t)}{t}&=\frac{1}{\epsilon}\paren{x_{b_i}(t) \paren{1-x_{b_i}(t)}+y_{b_i}(t)\paren{q-x_{b_i}(t)}}\\
\pd{y_{b_i}(t)}{t}&=\frac{1}{\epsilon'}\paren{2h z_{b_i}(t)-y_{b_i}(t)\paren{q+x_{b_i}(t)}}\\
\pd{z_{b_i}(t)}{t}&=x_{b_i}(t)-z_{b_i}(t) \\
\pd{x_{s_i}(t)}{t}&=\frac{1}{\epsilon}\paren{x_{s_i}(t) \paren{1-x_{s_i}(t)}+y_{s_i}(t)\paren{q -x_{s_i}(t)}+ \rho K_{\text{ex}}\paren{x_{b_i}(t)-x_{s_i}(t)}}\\
\pd{y_{s_i}(t)}{t} &=\frac{1}{\epsilon'}\paren{-y_{s_i}(t)\paren{q +x_{s_i}(t)}+\rho K_{\text{ex}}\paren{y_{b_i}(t)-y_{s_i}(t)}} \\
\end{aligned}
\right.
\end{equation} 
\end{widetext}
with $i=1,2$. For the set of parameters in the oscillatory regime as described above, the beads and the surrounding solution of each copy oscillate with a natural frequency~$\omega_0$.

We now couple the two systems by coupling the beads of one system to the solution of the second system. Specifically, we couple the bead of system $i=1$ with the surrounding solution of system $i=2$ using
\begin{equation}
\mathbf{G_1}=\begin{bmatrix}
-\frac{1}{\epsilon}\paren{x_{b_1}-x_{s_2}}\\
-\frac{1}{\epsilon'}\paren{y_{b_1}-y_{s_2}}\\
0\\
0\\
0
\end{bmatrix}
\end{equation}
and with a coupling strength $K = \frac{K_{\text{ex}}}{2\paren{1+\rho}}$. This now allows to derive a phase description if the coupling strength~$K$ is small. For simplicity, we absorb the coupling strength into the coupling function~$H_i$ so that the phase of oscillator~$1$ evolves according to
\begin{equation}
\d{\theta_{1}(t)}{t}= \omega_0+H_1(\theta_{2}-\theta_{1}).
\end{equation}
\begin{figure}[htp]
 \centering
 \includegraphics[width=0.75\textwidth]{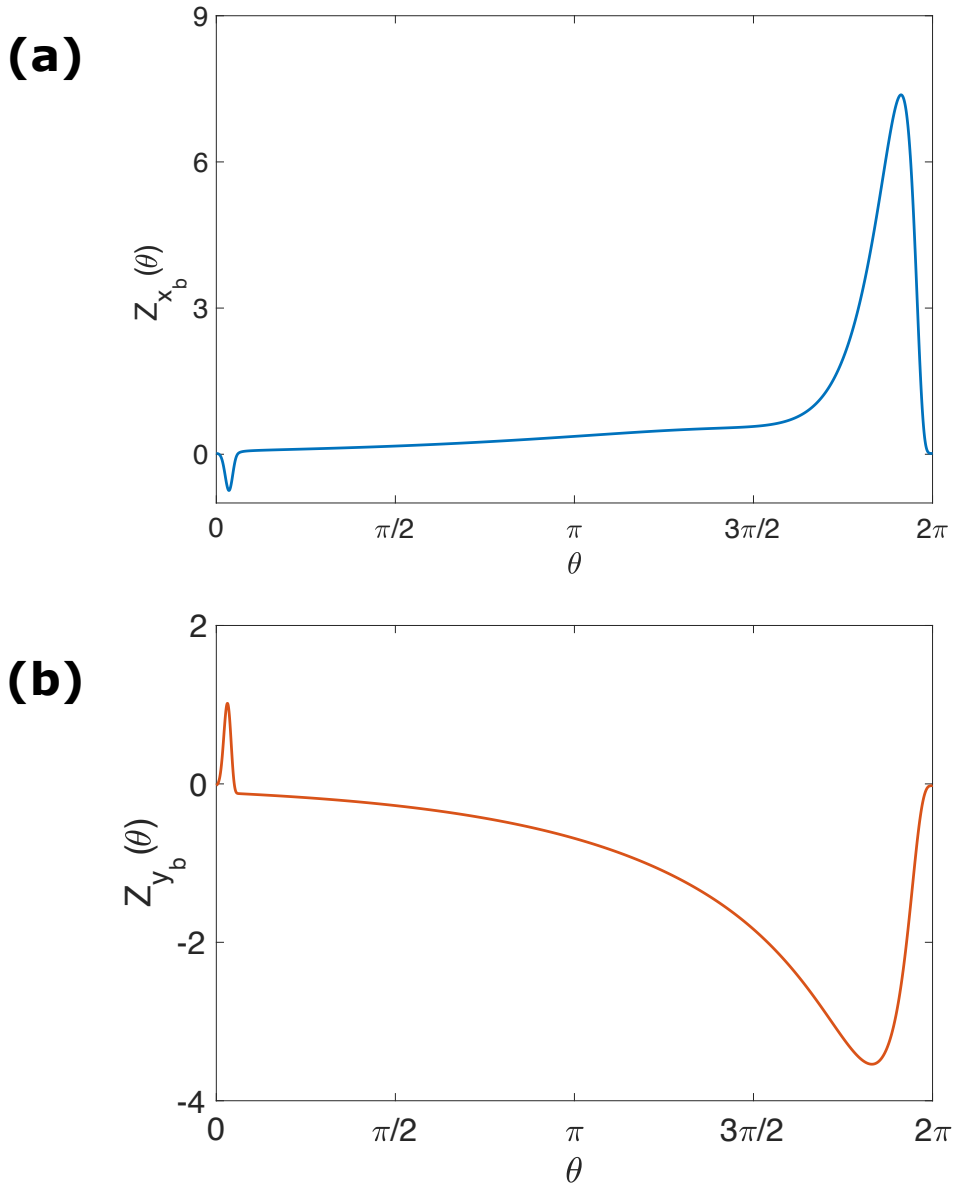}
   \caption{Phase response curves (PRC) for one bead with $\rho=1.2$. Chemical parameters set used: $q=0.002$ , $\epsilon=0.01$ , $\epsilon'=0.015$, $h=0.70$.}
 \label{Figure 3}
\end{figure}
Now, we can calculate the periods of the oscillators in both synchronized and super-synchronized states and compare the full nonlinear model and the phase approximation.
For the phase dynamics we know that the beads and the surrounding solution oscillate in phase, $\phi_0=0=2\pi$, with frequency~$\Omega$ and period $T=\frac{2\pi}{\Omega}$ so that
\begin{equation}
\label{eq:ph_eq}
\d{\theta_{1}(t)}{t}= \Omega= \frac{2\pi}{T}=\omega_0+H_1(0)
\end{equation}
For the full nonlinear system, the model chemical parameters were set 
\begin{equation}
q=0.002 \quad , \quad \epsilon=0.01 \quad , \quad \epsilon'=0.015 \quad , \quad h=0.70 
\end{equation}
The control parameters considered for the analysis are the chemical exchange rate~$K_{\text{ex}}$ and the density of beads in the medium~$\rho$.
The components of the PRC~$\mathbf{Z}(\theta)$) for one bead with a density $\rho=1.2$ are shown in \textbf{Figure~\ref{Figure 3}}.

\begin{figure}[htp]
 \centering
 \includegraphics[width=0.75\textwidth]{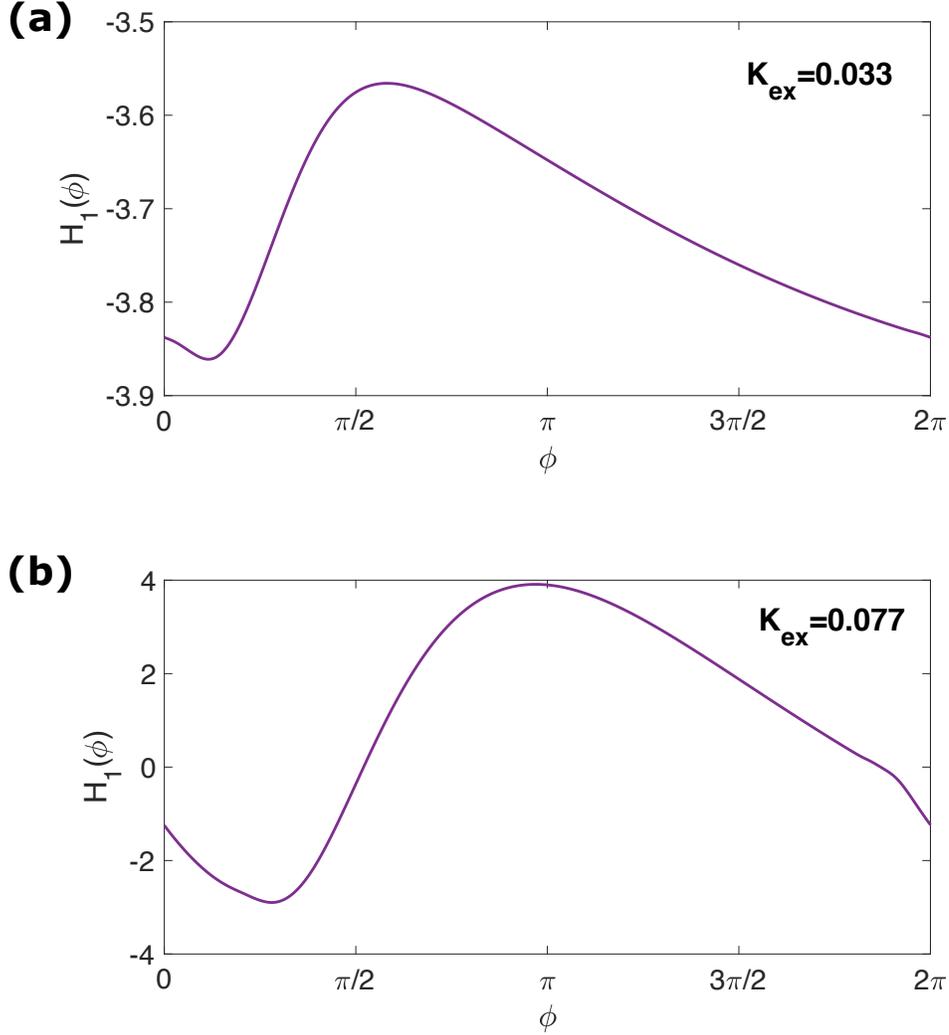}
   \caption{Two cases of $H_1(\phi)$ with density $\rho=1.2$ corresponding to (a) synchronized state with $K_{\text{ex}}=0.033$ and (b) super-synchronized state with $K_{\text{ex}}=0.077$. Note the remarkable difference in scale on the $y$-axis of both figures.}
 \label{Figure 4}
\end{figure}

Interaction functions ($H_1(\phi)$) in the synchronized and super-synchronized states and density $\rho=1.2$ are shown in \textbf{Figure~\ref{Figure 4}}.

\begin{figure}[htp]
 \centering
 \includegraphics[width=1.0\textwidth]{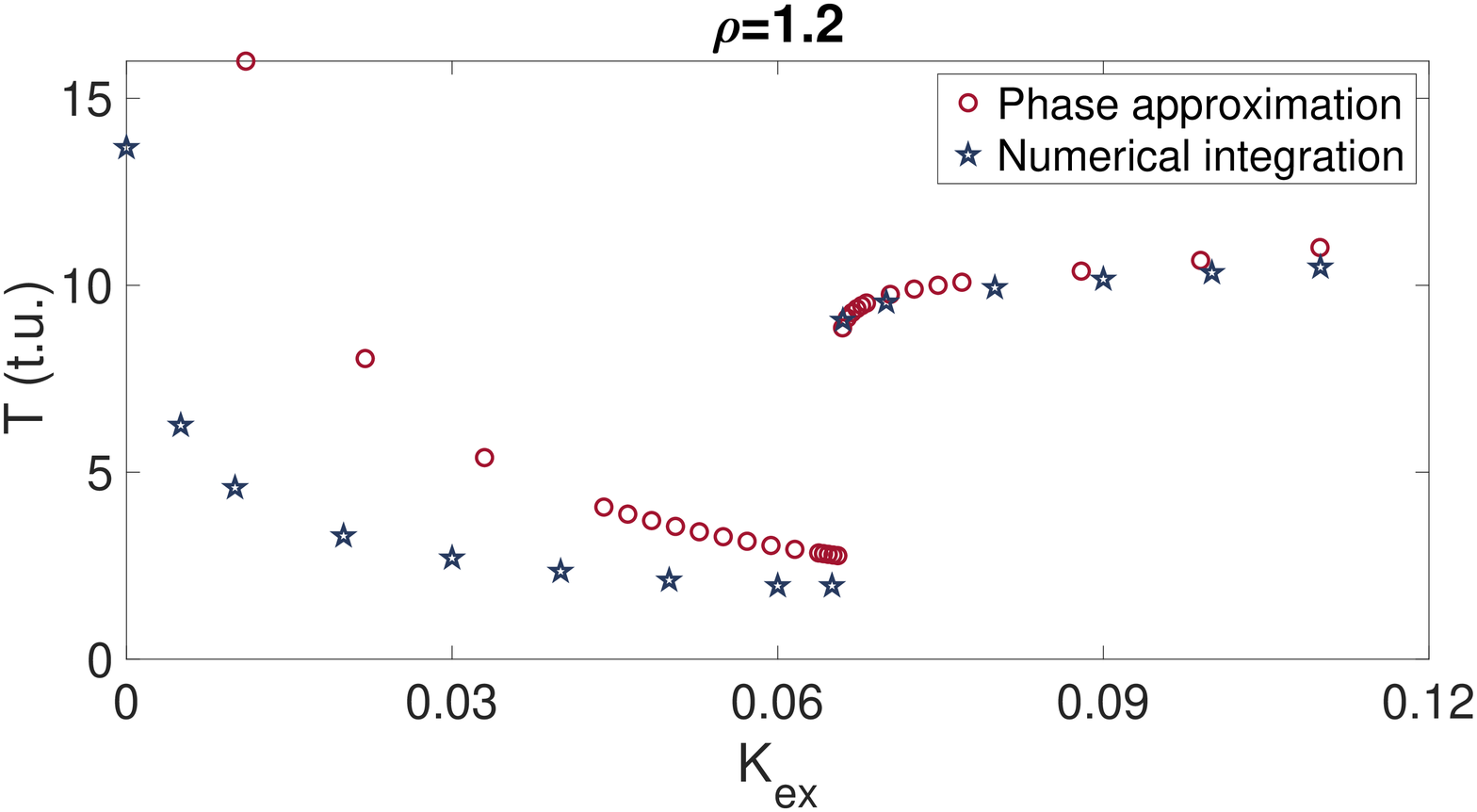}
   \caption{Periods of the oscillators in function of~$K_{\text{ex}}$ with density $\rho=1.2$. The discontinuity in period shows the transition from synchronized to super-synchronized states.}
 \label{Figure 5}
\end{figure}

The phase approximation yields a good description of the period of the collective oscillation as parameters are varied. The calculated periods for each value of~$K_{\text{ex}}$ obtained via numerical integration and by using the phase approximation model~\eqref{eq:ph_eq} with density $\rho=1.2$ are shown in \textbf{Figure~\ref{Figure 5}}. A transition from synchronized to super-synchronized phase is seen, with a discontinuity in the period and the interaction function caused by the bifurcation described between limit cycles of different nature.

\subsection*{Fourier expansion}

The qualitative change in the oscillation in the transition between the synchronized and super-synchronized states can also be seen in the change of interaction function of the phase reduction. Indeed, the changes in the phase interaction function are an indicator of the underlying bifurcations~\cite{Hesse2017}. To illustrate this effect in the chemical oscillator system, we expand the interaction function in Fourier series, and to understand how the Fourier modes change as the system parameters are varied.

We can expand the interaction function into a sine-cosine Fourier series (in the supplementary information a exponential Fourier series expansion is presented showing equivalent results),
\begin{equation}
H_1(\phi)=\frac{a_0}{2}+\sum_{k=1}^\infty \cor{a_k \cos \paren{k \phi}+b_k \sin \paren{k \phi}}
\end{equation} 

\begin{figure}[htp]
 \centering
 \includegraphics[width=0.75\textwidth]{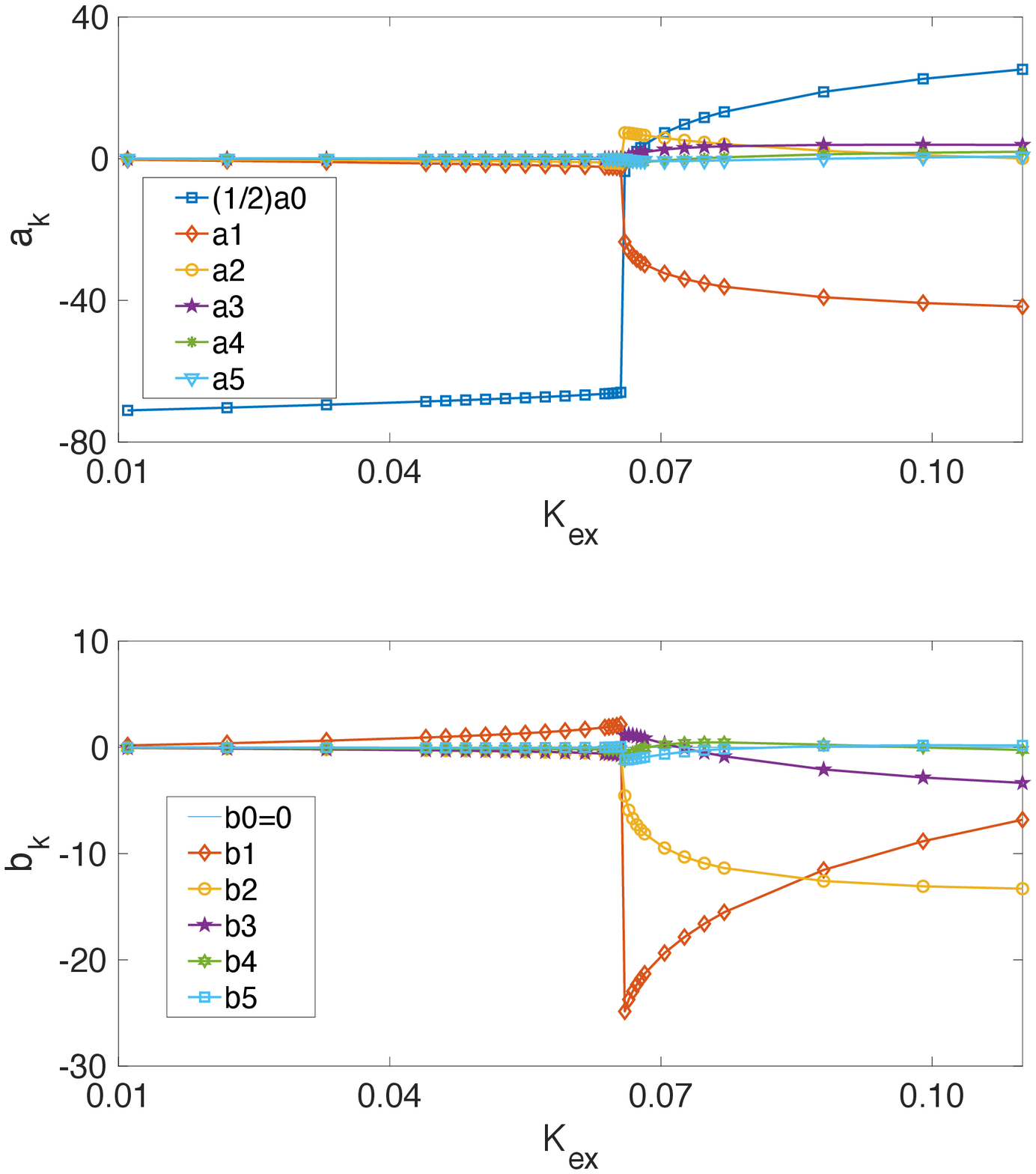}
   \caption{Fourier coefficients (sine-cosine serie). Due to the different nature of the limit cycles in the states of synchronization and super-synchronization, we have different modes in the function of interaction in these states.}
 \label{Figure 6}
\end{figure}

In \textbf{Figure~\ref{Figure 6}}, we show the coefficients obtained for each value of $K_{\text{ex}}$ ($H_1(\phi)$ calculated with the method referred above for each value of $K_{\text{ex}}$).
We can see that in the synchronized state:
\begin{equation}
H_1(\phi) \sim \frac{a_0}{2} + a_1 \cos \paren{\phi} + b_1 \sin \paren{\phi} \quad ,
\end{equation}
and the values of the coefficients hardly vary, except in the vicinity of the transition between synchronized and super-synchronized states.

On the other hand, in the super-synchronized state we have higher harmonics and the value of the coefficients varies considerably for a wide range of values of $K_{\text{ex}}$:
\begin{widetext}
\begin{equation}
H_1(\phi) \sim \frac{a_0}{2} + a_1 \cos \paren{\phi} + b_1 \sin \paren{\phi} + a_2 \cos \paren{2\phi} + b_2 \sin \paren{2\phi} + a_3 \cos \paren{3\phi} + b_3 \sin \paren{3\phi} 
\end{equation}
\end{widetext}

\section{Conclusions}

In this manuscript, we considered the problem of synchronization between oscillators embedded into an active medium. This type of system has shown to exhibit more than one state of synchronization. The chosen system is constituted by a set of chemical oscillators immersed into a chemical solution that provides the physical medium to interact but also adds dynamics to the total system. This system has been shown to synergetically produce a different synchronization state (supersynchronization) that was not accessible for each of the two main components of the problem (external medium or the beads).

Using continuation and bifurcation analysis, we reconstructed the experimental and numerical parameter space previously reported in \cite{Ghoshal2016}. Three different states of synchronization are found; oscillation death, synchronization and mobbing state (super-synchronization) and the transitions between each other analyzed from a bifurcation analysis point of view. Note that in the five-dimensional simplified  model used, the oscillation death correspondes with a steady state of the system, synchronization is a normal oscillatory behavior and supersynchronized state is demonstrated as a different state of oscillation.

In order to calculate the periods exhibited by the oscillators in each state, a phase approximation model was considered reproducing with good accuracy the previous results reported both in experiments and in numerical simulations as well as the discontinuity that signals the transition between synchronization to mobbing state.

Finally, we have proved  that the discontinuity in the periods and the interaction function in the transition from synchronized to super-synchronized states is caused by the bifurcation described between limit cycles of different nature. This was possible to understand considering that the active medium is an oscillator although with a different nature (the medium does not have a catalyst per se although the catalyst is included into the beads that are immersed in the medium and, thus, needs the activity of the beads to oscillate). 

The results of this analysis can extrapolate to different systems as far as the connective medium plays an active role in the dynamics of the system showing the generality of the phenomenon described. This type of system is found in different fields in Nature including neuronal processes involving the glia \cite{Alvarez-Maubecin2018}, as glial cells and neurons have ionic channels that allow them to oscillate but only neurons possess synaptic connections and have the ability to oscillate by themselves.

\begin{acknowledgments}
We gratefully acknowledge ﬁnancial support by the Spanish Ministerio de Econom\'{i}a y Competitividad and European Regional Development Fund under contract RTI2018-097063-B-I00 AEI/FEDER, UE, and by Xunta de Galicia under Research Grant No. 2018-PG082. Authors are part of the CRETUS Strategic Partnership (AGRUP2015/02). All these programs are co-funded by FEDER (UE).
\end{acknowledgments}

\bibliography{chem_osc_act_med_synch_PRE_bib}

\end{document}